\documentclass[%%aip,
 prd,
 jmp,%
 reprint,%
 onecolumn,
 superscriptaddress,
,12pt
]{revtex4-2}

\usepackage{amssymb}
\usepackage{amsmath}
\usepackage{graphicx}
\usepackage{dcolumn}
\usepackage{bm}
\usepackage{color}
\usepackage{xcolor}
\usepackage{float}
\allowdisplaybreaks
\usepackage{subfig}
\usepackage{comment}
\usepackage{multirow}
\usepackage[figurename=FIGURE,tablename=TABLE]{caption}
\usepackage{threeparttable}
\usepackage{hyperref}
\hypersetup{
    colorlinks=true, %set true if you want colored links
    linktoc=all,     %set to all if you want both sections and subsections linked
    linkcolor=magenta,
    citecolor=red
}

\begin{document}

\title{Non-minimal coupled warm inflation with quantum-corrected self-interacting inflaton potential}

\author{Daris Samart}
\email{darisa@kku.ac.th}
\affiliation{Khon Kaen Particle Physics and Cosmology Theory Group (KKPaCT), Department of Physics, Faculty of Science, Khon Kaen University, 123 Mitraphap Rd., Khon Kaen, 40002, Thailand}
 %\altaffiliation[Also at ]{Physics Department, XYZ University.}%Lines break automatically or can be forced with \\
\author{Patinya Ma-adlerd}%
\email{m.patinya@kkumail.com}
\affiliation{Khon Kaen Particle Physics and Cosmology Theory Group (KKPaCT), Department of Physics, Faculty of Science, Khon Kaen University, 123 Mitraphap Rd., Khon Kaen, 40002, Thailand}

\author{Peeravit Koad}
\email{harrykoad@gmail.com}
\affiliation{School of Informatics, Walailak University, Thasala, Nakhon Si Thammarat, 80160, Thailand}

\author{Phongpichit Channuie}
\email{channuie@gmail.com}
\affiliation{College of Graduate Studies, Walailak University, Thasala, Nakhon Si Thammarat, 80160, Thailand}
\affiliation{School of Science, Walailak University, Thasala, Nakhon Si Thammarat, 80160, Thailand}

%\author{C. Author}
% \homepage{http://www.Second.institution.edu/~Charlie.Author.}
%\affiliation{%
%Second institution and/or address%\\This line break forced% with \\
%}%

\date{\today}% It is always \today, today,
             %  but any date may be explicitly specified

\begin{abstract}
In this work, we investigate the non-minimally coupled scenario in the context of warm inflation with quantum-corrected self-interacting potential. We transform the potential in the Jordan frame to the Einstein frame and consider a dissipation parameter of the form $\Gamma = C_{T}T$ with $C_T$ being a coupling parameter. We focus on the strong regime of which the interaction between inflaton and radiation fluid has been taken into account. We compute inflationary observables and constrain the parameters of our model using current Planck 2018 data. With the sizeable number of {\it e}-folds and proper choices of parameters, we discover that allowed values of $C_T$ lie in the range $0.014\lesssim C_{T}\lesssim 0.020$ in which the predictions are in good agreement with the latest Planck 2018 results at the $2\,\sigma$ confident level.
\end{abstract}

\maketitle

%\tableofcontents
\newpage
%%%%%%%%%%%%%%%%%%%%%%%%%%%%
\section{Introduction}
%%%%%%%%%%%%%%%%%%%%%%%%%%%
%Starting with the warm inflation
Warm inflation scenario has been received attentions as an alternative approach of the reheating phase of the universe in order to generate the thermal bath in the standard cosmology. The warm inflation was originally proposed to resolve some problems in the standard cold inflation picture \cite{Berera:1995ie,Berera:1995wh}, for instances, providing sufficiently hot thermal bath after inflaton decaying to other matter fields in the reheating epoch \cite{Berera:1996nv,Hall:2003zp} the large quantum correction of the inflaton field might spoiling the flatness of the observed universe or a so-called eta problem \cite{Berera:1999ws,Berera:2003yyp}, fine tuning of the initial values of the inflation models motivated by beyond standard model physics \cite{Ramos:2001zw,Berera:2000xz,Bastero-Gil:2016mrl} and other salient features see \cite{Berera:2008ar,Bastero-Gil:2009sdq,Rangarajan:2018tte} for reviews. In the warm inflation stage, the inflaton decays into radiation matter during the slow-roll period. In the meantime, the quantum fluctuations of the density perturbation amplitudes are generated by the friction of the inflaton propagating in the thermal bath. At the end of inflation, the universe is automatically heated up with out requiring the preheating and reheating phases before radiation dominated era. Moreover, the energy density of the radiation is smoothly joined with the energy density of the inflaton field. The dissipative coefficient, $\Gamma$ plays a crucial role in the warm inflationary universe for describing the dynamics of warm inflation. All information of the microscopic dynamical processes during warm inflationary universe is contained in the dissipative coefficient and it has been constructed and calculated by using the supersymmetric models with finite temperature field analysis in various aspects see Refs. \cite{Moss:2006gt,Berera:1998gx,Berera:2001gs,Zhang:2009ge,Bastero-Gil:2011rva,Bastero-Gil:2012akf,Bastero-Gil:2014oga,Berera:2002sp,Bastero-Gil:2005klw,Bastero-Gil:2004oun} for more details and references therein.  

%talk about the quantum correction in self-interacting potential
The self-interacting inflaton potential ($V \sim \phi^4$) has been largely used to study of the standard (cold) inflation dynamics in numerous perspectives see \cite{Bassett:2005xm,Martin:2013tda,Gron:2018rtj} for reviews. According to the requirements of the standard quantum field theory, the self-interacting potential is renormalizable theory and it is naturally received the quantum-corrected effect. The quantum correction of the perturbative loop expansion known as Coleman-Weinberg potential \cite{Coleman:1973jx} is one of the famous approach. In addition, the phenomenological quantum-corrected self-interacting potential is proposed and employed to study the quantum-corrected effect due to the non-vanishing primordial tensor modes by Ref.\cite{Joergensen:2014rya}. On the other hand, there are a number of investigations that also used the self-interacting potential to study warm inflationary universe in both minimal and non-minimal coupling to gravity, for examples see Refs.\cite{Panotopoulos:2015qwa,Benetti:2016jhf,Motaharfar:2018mni,Kamali:2018ylz,Graef:2018ulg,Arya:2018sgw,Bastero-Gil:2018uep}. However, there is no previous work in a study of non-minimally coupled warm inflation with the quantum-corrected self-interacting potential according to the literature. 

%state the aim of this paper and organize %temperature manifest
Therefore, in this work, we will investigate the quantum-correction of the self-interacting potential due to the thermal effect inflation with $\Gamma \propto T$, where $T$ is a temperature. Our study might shed some light on the the quantum-correction of the inflaton due to the finite temperature reaction which plays significant role in warm inflationary universe. In particular, the results in this work might reveal to what extend the model's parameters deviate from cold inflation when the thermal effect is taken into account. Moreover, we will constrain our theoretical results with Planck 2018 via the COBE normalization and the prediction in this work will be compared to the latest observational data.   

The paper is organized as follows: all relevant dynamical equations in the non-minimal coupling warm inflation under the slow-roll approximation are determined in section \ref{formalism}. Next, in section \ref{confront-data}, we will compare the results in this work with the observational data. Finally, we close this paper by providing discussions and conclusions in section \ref{conclusion}.  

%%%%%%%%%%%%%%%%%%%%%%%%%%%%
\section{Formalism} 
\label{formalism}
%%%%%%%%%%%%%%%%%%%%%%%%%%% 

\subsection{Non-minimal coupling gravitational action and conformal transformation}
We start with a gravitational action of the non-minimal coupling of the scalar field to Ricci scalar (gravity) with a general form of the effective potential $V(\phi)$, one finds,
\begin{eqnarray}
S_J = \int \sqrt{-g}\left[ -\frac12\left(M_p^2 + \xi \,\phi^2 \right)R + g^{\mu\nu}\partial_\mu\phi\partial_\nu\phi - V(\phi)\right]
\label{action-J}
\end{eqnarray}
where the action $S_J$ stands for the gravitational action in the Jordan frame. While $M_p^2\equiv 1/8\pi G$ and $\xi$ are reduced Plank mass and the non-minimal coupling constant, respectively. It is more convenient to study the inflation dynamics of the non-minimal coupling in the Einstein frame, i.e., the gravitational sector of the action written in the Einstein-Hilbert form only. The Einstein frame can be achieved by using the conformal transformation via a re-defining metric tensor as,
\begin{eqnarray}
\widetilde{g}_{\mu\nu} = \Omega(\phi)^2\,g_{\mu\nu}\,, \qquad \Omega(\phi) = 1+ \frac{\xi\,\phi^2}{M_p^2}\,.
\end{eqnarray}
Here all variables with tilde symbol represent the quantities in the Einstein frame. Applying the conformal transformation to the action (\ref{action-J}), the action in Einstein frame is given by,
\begin{eqnarray}
S_E = \int \sqrt{-\widetilde{g}}\left[ -\frac12\,M_p^2\,\widetilde{g}^{\mu\nu}\widetilde{R}_{\mu\nu} + \widetilde{g}^{\mu\nu}\partial_\mu\chi\partial_\nu\chi - U(\chi)\right].
\label{action-E}
\end{eqnarray}
We have used the re-definition of new scalar field, $\chi$ in the Einstein frame to obtain the canonical form of the kinetic term of the scalar field, $\chi$ as
\begin{eqnarray}
\frac12\left( \frac{d\chi}{d\phi}\right)^2 = \frac{1+ 3\,M_p^2\,\Omega_\phi^2}{\Omega^2} \,,
\end{eqnarray}
where $\Omega_\phi\equiv d \Omega/d\phi$ and the new effective potential in the Einstein frame, $U(\chi)$ is also given by,
\begin{eqnarray}
U(\chi) = \Omega^{-4}\,V\left( \phi(\chi)\right).
\label{E-potential}
\end{eqnarray}
In this work, we will consider the self-interacting potential with phenomenological quantum correction in the warm inflation scenario. This potential has been proposed by Ref.\cite{Joergensen:2014rya} in order to analyze the characters of the quantum correction in the self-interacting scalar field phenomenologically. The potential in the Jordan frame is written in the following form
\begin{eqnarray}
V(\phi) = \lambda\,\phi^4\left( \frac{\phi}{\Lambda}\right)^{4\gamma}\,.
\label{quantum-potential}
\end{eqnarray}
Here we introduced the quantum correction (real) parameter $\gamma$ that use to characterize the quantum behavior of the self-interacting potential and the $\Lambda$ parameter is the cut-off at a given energy scale. It was shown that the range of the $\gamma$ should be $\mathcal{O}(\gamma) \sim 0.1$ according to the constraint from observational data \cite{Joergensen:2014rya}. In the latter, we will construct the slow-roll dynamics in warm inflation in the Einstein frame with the potential in Eq.(\ref{quantum-potential}). 

%%%%%%%%%%%%%%%%%%%%%%%%%%%%%%%%%%%%%%%%%%%%%%%%%%%%%%%%%%%%%%%%%%%%%%%%%%%%
\subsection{Slow-roll dynamics in warm inflation}
%%%%%%%%%%%%%%%%%%%%%%%%%%%%%%%%%%%%%%%%%%%%%%%%%%%%%%%%%%%%%%%%%%%%%%%%%%%%
In this subsection, we collect all relevant cosmological equations of the slow-roll paradigm in warm inflation. Recalling the Friedmann equation from the gravitational action in Einstein frame with the flat FRW background in the warm inflation scenario, it reads,
\begin{eqnarray}
H^2 = \frac{1}{3\,M_p^2}\left( \frac{1}{2}\,\dot\chi^2 + U(\chi) + \rho_R\right),
\end{eqnarray}
where $\dot\chi\equiv d\chi/dt$ and $\rho_R$ is energy density of the radiation fluid with the equation of state $w_R=1/3$. The Klein-Gordon equation of motion for the scalar field, $\chi$ in Einstein frame with the dissipative coefficient, $\Gamma$ is given by
\begin{eqnarray}
\ddot\chi + 3H\,\dot\chi + U_\chi = -\Gamma\,\dot\chi\,,
\end{eqnarray}
with $U_\phi \equiv dU/d\chi$. The conservation of the energy-momentum tensor of the radiation fluid leads to the continued equation as
\begin{eqnarray}
\dot\rho_R + 4H\,\rho_R = \Gamma\,\dot\chi^2\,.
\end{eqnarray}
Based on the finite temperature field theory approaching to the supersymmetry models, the generic form of the dissipative coefficient, $\Gamma$ for several warm inflation models can be written in terms of the coupling between temperature $(T)$, scalar (inflaton) field $(\chi)$ and the mass of the some heavy field during warm inflation $(M_X)$ with $M_X>T$ as the following form \cite{Berera:1998gx,Berera:2001gs,Zhang:2009ge,Bastero-Gil:2011rva,Ramos:2013nsa}
\begin{eqnarray}
\Gamma = C_{(m)}\,\frac{T^m\,\chi^{n}}{M_X^l}\,,
\qquad m +n-l=1\,.
\label{general-gamma}
\end{eqnarray}
The dissipative coefficient, $\Gamma$ represents the energy transfer from the inflaton field to the thermal bath in the warm inflationary universe. The parameter $C_m$ encodes the microscopic dynamics of the inflaton interacting with other particles and the $m$, $n$ and $l$ are the integer number. Particularly for the $m=1$ case, this corresponds to the high temperature supersymmetric model \cite{Zhang:2009ge} or considering inflaton as a pseudo Goldstone boson that can be coupled to other fields in the thermal bath such as warm natural inflation \cite{Mishra:2011vh} and warm little inflation in analogy to the little Higgs model \cite{Bastero-Gil:2016qru}.  % On the other hand, with $m=3$ and $n=-2$ case implies the low energy approximation of the supersymmetric thermal field theory for the super scalar (inflaton) field couples to the intermediate heavy field where the mass of the heavy field is larger than the raidation temperature. Therefore the inflaton decouples to the radiation field and this leads to a large dissipation and heavy field through the warm inflation \cite{Bastero-Gil:2010dgy,Bastero-Gil:2012akf}. 
In the following, we will consider the slow-roll approximation framework with the dissipative coefficients $\Gamma$ for $m=1$ and at the strong regime.

In the standard slow-roll approximation, we can re-write the Friedmann equation as well as the equations of motion for the inflaton and the radiation matter as
\begin{eqnarray}
H^2 &\approx& \frac{1}{3 M_p^2}\,U(\chi)\,,
\label{SR-friedmann}
\\
\dot\chi &\approx& -\frac{U_\chi}{3H(1+Q)}\,,\qquad Q\equiv \frac{\Gamma}{3H}\,,
\label{SR-KG}
\\
\rho_R &\approx& \frac{\Gamma\,\dot\chi}{4H}\,,\qquad \rho_R = C_R\,T^4\,,
\label{SR-rad}
\end{eqnarray}
where the $Q$ is called dimensionless parameter that use to identify the regime of the dissipative effects in the latter and $C_R = g_R\,\pi^2/30$. In addition, the minimal supersymmetric standard model gives the number of relativistic degrees of freedom, $g_R = 228.76$ and leads to $C_R \simeq 70$ \cite{Hall:2003zp}. We have been used the following approximations for the slow-roll scenario,
\begin{eqnarray}
\rho_R &\ll& \rho_\chi\,,\qquad 
\rho_\chi = \frac12\,\dot\chi^2 + U\,,
\\
\dot\chi^2 &\ll& U(\chi)\,,
\\
\ddot\chi &\ll& 3H\left( 1 + Q\right)\dot\chi\,,
\\
\dot\rho_R &\ll& 4H\,\rho_R\,,
\end{eqnarray} 
As mentioned earlier, it is more convenient to separate warm inflation into two regimes by using the dimensionless $Q$ as
\begin{eqnarray}
Q &\gg& 1\,,\qquad {\rm strong~regime}\,,
\\
Q &\ll& 1\,,\qquad {\rm weak~regime}\,.
\end{eqnarray}
In addition, we can express the temperature as a function of the inflaton field, $\chi$ by using the Eqs. (\ref{general-gamma},\ref{SR-friedmann},\ref{SR-KG},\ref{SR-rad}) for the general $m$ integer values. It reads,
\begin{eqnarray}
T &=& \left( \frac{U_\chi^{2}\,\chi^{m-1}}{4H\,C_{(m)}\,C_R}\right)^{\frac{1}{4+m}}\,,\qquad {\rm for}\,~Q\gg 1\,,
\label{T-phi-strong}
\\
T &=& \left( \frac{C_{(m)}\,U_\chi^{2}\,\chi^{1-m}}{36H^3\,C_R}\right)^{\frac{1}{4-m}}\,,\qquad {\rm for}\,~Q\ll 1\,.
\label{T-phi-weak}
\end{eqnarray}

In the following, we will concentrate and investigate warm inflation in the strong dissipative regime only since this regime might show better thermal effect in the inflationary universe.   

Before calculating the slow-roll parameters, we would like to express the form of the effective potential in the Einstein in Eq.(\ref{E-potential}) under the large field assumption during the inflation i.e., $ \phi \gg {M_p /\sqrt{\xi}}$\,. One finds,
\begin{eqnarray}
\chi \simeq \kappa\, M_p \ln \Big({ \sqrt{\xi}\, \phi \over  M_p } \Big) , ~~~~ \kappa \equiv \sqrt{{2\over \xi} + 6} 
\end{eqnarray}
Then the Einstein frame potential then takes the following form 
\begin{eqnarray}
U(\chi) = \Omega^{-4} V(\phi(\chi)) &=& \frac{ M_p^4}{\left(M_p^2+\xi  \phi ^2\right)^2} \lambda\,  \phi ^4  \left(\frac{\phi }{\Lambda }\right)^{4 \gamma }
\nonumber\\
&=& {\lambda M_p^4 \over \xi^2} \left(\exp\Bigg[{\frac{-2 \chi}{\kappa M_p}} \Bigg] +1 \right)^{-2}  \left({M_p \over \sqrt{\xi} \Lambda} \right)^{4 \gamma} \exp \Bigg[{4 \gamma \chi \over \kappa M_p} \Bigg] 
\label{U-chi}
\end{eqnarray} 
Next, we provide the slow-roll parameters in warm inflation for general $m$ and they read,
\begin{eqnarray}
\epsilon &=& \frac{M_p^2}{2}\left( \frac{U_\chi}{U}\right)^2\,,\quad \eta = M_p^2\,\frac{U_{\chi\chi}}{U}\,,\quad \beta = M_p^2\left( \frac{U_\chi\,\Gamma_\chi}{U\,\Gamma}\right)\,.
\label{SR-parameters}
\end{eqnarray}
The inflationary phase of the universe occurs under the following conditions
\begin{eqnarray}
\epsilon \ll 1 + Q\,,\qquad \eta \ll 1 + Q\,,\qquad \beta \ll 1 + Q\,.
\end{eqnarray}
We firstly calculate the slow-roll parameters, $\epsilon$ and $\eta$ with the potential in Eq.(\ref{E-potential}) and they are given by
\begin{eqnarray}
\epsilon &\approx& %\frac{8 M_p^4}{\kappa ^2 \xi ^2 \phi ^4} +\frac{16\,M_p^2}{\kappa^2 \xi  \phi ^2}\,\gamma+\frac{8 }{\kappa^2}\,\gamma ^2 =
\frac{8}{\kappa^2}\left(\frac{M_p^2}{\xi\,\phi^2}\right)^2\left(1 + \gamma\,\frac{\xi\,\phi^2}{M_p^2} \right)^2
\label{SR-epsilon}
\\
\eta &\approx&  %\frac{16 M_p^4}{\kappa^2 \xi ^2 \phi ^4}-\frac{8 M_p^2}{\kappa^2 \xi  \phi ^2} + \frac{32 M_p^2}{\kappa^2 \xi  \phi ^2}\, \gamma +\frac{16 }{\kappa^2}\,\gamma ^2 
\frac{16}{\kappa^2}\,\frac{M_p^4}{\xi^2\,\phi^4} \left(1 + \left(2 \gamma -\frac12\right)\frac{\xi\,\phi^2}{M_p^2}\right)
\,,
\label{SR-eta}
\end{eqnarray}
where we have presented $\chi$ of the Einstein frame in terms of $\phi$ of the Jordan frame with $\phi = M_p\,\exp\left({{\chi}/{\kappa\,M_p}}\right)/\sqrt{\xi}$\,.
The $\beta$ parameter depends on the dissipative coefficient, $\Gamma$\, and we can calculate the $\beta$ parameter after introducing the explicit form of the $\Gamma$. 

Next We start with the dissipative coefficient of warm inflation, the dissipative coefficient for $m=1$ is read,
\begin{eqnarray}
\Gamma = C_{T}\,T\,.
\label{C1}
\end{eqnarray}
On the one hand, the dissipative coefficient in this form can be achieved from high temperature approximation of the thermal supersymmetric model \cite{Moss:2006gt}. On the other hand, the warm little inflation \cite{Bastero-Gil:2016qru} might generates the disipative coefficient in (\ref{C1}) where the inflaton in this scenario is considered as a pseudo Nambu-Goldstone boson of a broken gauge symmetry in the warm little inflation similar to "Little Higgs" model for electroweak symmetry breaking. Moreover, warm inflation can naturally occur for $T>H$. The coupling $C_T$ in Eq.(\ref{C1}) is given by 
\begin{eqnarray}
C_T \simeq \frac{3\,g^2}{h^2\big(1-0.34\log(h) \big)}\,,
\label{C11}
\end{eqnarray}
where $g$ is the Yukawa coupling of the inflaton (super scalar field) and heavy fermions in the warm little inflation scenario while $h$ is Yukawa coupling of the heavy fermions and light singlet scalar and fermion fields \cite{Bastero-Gil:2016qru}. 

By using the dissipative coefficient in Eq.(\ref{C1}), this leads to the expression of the temperature as a function of the inflaton field by using Eq.(\ref{T-phi-strong}) as
\begin{eqnarray}
T = \left( \frac{U_\chi^{2}}{4H\,C_T\,C_R}\right)^{\frac{1}{5}}.
\label{T1}
\end{eqnarray}
Having use the results in Eqs.(\ref{U-chi}), (\ref{C1}) and (\ref{T1}), the slow-roll parameter $\beta$ up to the first order of the $\gamma$ correction at the large field approximation is given by
\begin{eqnarray}
\beta \approx \frac{24\,M_p^4}{5\,\kappa^2\,\xi^2\,\phi^4} - \frac{16\,M_p^2}{5\,\kappa^2\,\xi\,\phi^2} + \frac{8\,M_p^4}{\kappa^2\,\xi\,\phi^2}\,\gamma\,,
\label{beta-T1}
\end{eqnarray}
where the relation $\phi = M_p\,\exp\left({{\chi}/{\kappa\,M_p}}\right)/\sqrt{\xi}$\, is implied. In addition, the dimensionless parameter $Q$ for $\Gamma = C_T\,T$ can be written by
\begin{eqnarray}
Q = \left[\left(\frac{2}{3}\right)^{2}\mathcal{K}\,\frac{M_p^4}{\phi^{4}}
\left(\frac{\Lambda}{\phi}\right)^{4\,\gamma}\left(1+ \gamma\,\frac{\xi  \phi ^2}{M_p^2}  \right)^{2}\,\right]^{\frac15}\,,\qquad \mathcal{K}\equiv \frac{ C_T^{4}}{C_R\,\kappa^2\,\lambda}\,.
\label{Q-T1}
\end{eqnarray}
At the end of inflation requiring $\epsilon_{\rm end} = Q$, one finds
\begin{eqnarray}
\frac{8}{\kappa^2}\left[\frac{M_p^4}{\xi^2\,\phi_{\rm end}^4}\left(1+ \gamma\,\frac{\xi  \phi_{\rm end} ^2}{M_p^2}  \right)^2\,\right]^{\frac45} 
&=& \left[\left(\frac{2}{3}\right)^{2}\mathcal{K}\,\xi^2
\left(\frac{\Lambda}{\phi_{\rm end}}\right)^{4\,\gamma}\right]^{\frac15}
\nonumber\\
\left(1+ \gamma\,\frac{\xi  \phi_{\rm end} ^2}{M_p^2}  \right)
&=&
\widetilde{K}\,\Lambda^2\,\frac{\xi}{M_p^2}\left(\frac{\phi_{\rm end}}{\Lambda} \right)^{2-\frac{\gamma}{2}}\,,
\label{epsilon-end-C1}
\end{eqnarray}
where the $\widetilde{K}$ parameter is defined by
\begin{eqnarray}
\widetilde{K} \equiv \left[ \left( \frac{\kappa^2}{8}\right)^5\,\left(\frac{2}{3}\right)^{2}\,\mathcal{K}\,\xi^2\,\right]^{\frac18}\,.
\label{def-Ktilde}
\end{eqnarray}
Applying the assumption that the given order of the $\gamma$ parameter, $\mathcal{O}(\gamma)\lesssim 0.1$ as mentioned earlier and leading to $2-\gamma/2 \approx 2$, the inflaton field at the end of warm inflation is read
\begin{eqnarray}
\phi_{\rm end} = \frac{1}{\sqrt{\widetilde{K} - \gamma}}\,\frac{M_p}{\sqrt{\xi}}
\label{phi-end-C1}
\end{eqnarray}
Therefore, the universal bound of the quantum correction for the self-interacting inflaton field due to the modification of warm inflation is given by
\begin{eqnarray}
\gamma < \widetilde{K}\,.
\label{gamma-bound-C1}
\end{eqnarray}
The bound in Eq.(\ref{gamma-bound-C1}) represents the thermal effects on the quantum-corrected parameter, $\gamma$ in warm inflation. It is worth noting that the universal bound in warm inflation is different from the standard (cold) inflation given by Ref.\cite{Joergensen:2014rya} as $\gamma < \sqrt{3}/2$ which is equivalent to the weak regime of warm inflation, i.e., $\epsilon(\phi_{\rm end})=1$. For given values $C_T = 0.02$, $C_R = 70$, $\xi = 10^4$ and $\lambda = 0.5\times10^{-4}$, we find $\widetilde{K} = 1.106$. This means the inflaton value at the end of warm inflation is smaller than that of the inflaton in cold inflation. 
\\
Moreover, the e-folding number, $N$ in the strong regime $Q\gg 1$ is given by
\begin{eqnarray}
N &=&\frac{1}{M_p^2} \int_{\chi_{\rm end}}^{\chi_N}\frac{Q\,U}{U_\chi}\,d\chi 
\nonumber\\
&=& \int_{\phi_{\rm end}}^{\phi_N}\frac{Q(\phi)}{\sqrt{2\,\epsilon(\phi)}}\,\frac{1}{\phi}\,d\phi
\approx \frac{5\,\kappa\,\xi}{24}\left[\frac{\mathcal{K}}{9\sqrt{2}}\left(\frac{\Lambda}{M_p}\right)^6\left(\frac{\phi}{\Lambda}\right)^{6\left(1-\frac{2}{3}\, \gamma\right)}\,\right]^{\frac15}\Bigg|_{\phi_{\rm end}}^{\phi_N}\,,
\end{eqnarray}
where we have expanded the $\gamma$ parameter up to the first order. Having use the $\phi_N\gg\phi_{\rm end}$ and $\mathcal{O}(\gamma)\sim 0.1$ approximations, the inflaton field at the Hubble horizon crossing in terms of the {\it e}-folding number, $N$ is written by
\begin{eqnarray}
\phi_{N} &\approx& 
\left[\frac{2187}{3125}\cdot 32768 \sqrt{2}\right]^{\frac{1}{6-4 \gamma }}\left[\left( \frac{N}{\kappa\,\xi}\right)^5
\left(\frac{M_p^6}{\mathcal{K}\,\Lambda^{4\gamma}} \right)\right]^{\frac{1}{6-4 \gamma}}
\nonumber\\
&=& \frac{12\cdot 2^{\frac{7}{12}}\,3^{\frac16}}{5^{\frac56}}
\left( \frac{N}{\kappa\,\xi}\right)^{\frac56}
\left(\frac{M_p}{\mathcal{K}^{\frac16}} \right)
\nonumber\\
&&\times\,\Bigg[1 +\frac{\gamma}{18}\left(\ln\left[\frac{2^{31}\cdot 3^{14}}{5^{10}}\right] + 2 \ln \left[\left( \frac{N}{\kappa\,\xi}\right)^5
\left(\frac{M_p^6}{\mathcal{K}\,\Lambda^6} \right)\right] 
\right)+\mathcal{O}\left(\gamma ^2\right)\Bigg]
\nonumber\\
&\simeq& 5.647
\left[ \frac{N^5}{\kappa^5\,\xi^2\,\mathcal{K}}\right]^{\frac16}
\Bigg[1 +\frac{\gamma}{9}\left(10.387 + \ln \left[\left( \frac{N}{\kappa\,\xi}\right)^5
\left(\frac{M_p^6}{\mathcal{K}\,\Lambda^6} \right)\right] \right)\Bigg]\frac{M_p}{\sqrt{\xi}} \,,
\label{phi-N}
\end{eqnarray}
here we have considered the quantum-corrected character of the self-interacting potential in warm inflation up to the leading order of the $\gamma$ parameter only. In addition, one might re-write the dimensionless parameter $Q$ in Eq.(\ref{Q-T1}) in terms of the {\it e}-folding number, $N$ by using Eq.(\ref{phi-N}) as
\allowdisplaybreaks
\begin{eqnarray}
Q(N) &\approx& \left[\left(\frac{2}{3}\right)^{2}\frac{\mathcal{K}\,M_p^4}{\phi_N^{4}}\left(\frac{\xi^2\,\Lambda^4}{M_p^4}  \right)^{\gamma}\,\right]^{\frac15} 
\nonumber\\
&\simeq& \left[\left(\frac{2}{3}\right)^{2}\frac{\mathcal{K}\,M_p^4}{\phi_N^{4}}\left(1 + 2\,\gamma\,\ln\left[\frac{\xi\,\Lambda^2}{M_p^2}\right]  \right)\right]^{\frac15} 
\nonumber\\
&=& \left[
\frac{\left(\frac{2}{3}\right)^{2}\frac{\mathcal{K}^{\frac53}\,\xi^2}{5.647^4}\left(1 + 2\,\gamma\,\ln\left[\frac{\xi\,\Lambda^2}{M_p^2}\right]  \right)}{\left[ \frac{N^5}{\kappa^5\,\xi^2}\right]^{\frac{2}{3}}
\left[1 +\frac{\gamma}{9}\left(10.387 + \ln \left[\left( \frac{N}{\kappa\,\xi}\right)^5
\left(\frac{M_p^6}{\mathcal{K}\,\Lambda^6} \right)\right] \right)\right]^{4}}\right]^{\frac15}
\label{Q-N}
\end{eqnarray}

Taking the back reaction of the inflaton fluctuation in the thermal heat bath into account, in addition, the power spectrum is given by \cite{Bastero-Gil:2011rva,Graham2009,Bastero-Gil:2018uep,Hall:2003zp,Ramos:2013nsa,Bastero-Gil:2009sdq,Taylor:2000ze,DeOliveira:2001he},
\begin{eqnarray}
\Delta_{\mathcal{R}} &=&  \frac{U\,\big(1 + Q_N\big)^2}{24\,\pi^2\,M_p^4\,\epsilon}\left(1 + 2\,n_N + \left(\frac{T_N}{H_N}\right)\frac{2\sqrt{3}\,\pi\,Q_N}{\sqrt{3+4\pi\,Q_N}}\right)G(Q_N)\,,
\nonumber\\
&\simeq& \frac{5\,C_T^3}{12\,\pi^4\,g_R\,Q_N^2} \left(1+ \frac{\sqrt{3}\,\pi\,Q_N}{\sqrt{3+4\pi\,Q_N}}\right)G(Q_N)\,
\label{spectrum}
\end{eqnarray}
where the subscript $``N"$ is labeled for the values of all quantities in warm inflation at the Hubble horizon crossing and $n = 1/\big( \exp{H/T} - 1 \big)$ is the Bose-Einstein statistical function. In addition, the power spectrum in Eq.(\ref{spectrum}) can be constrained by observational data and yields the upper bound of the $C_T$ parameter as $C_T \lesssim 0.02$ \cite{Bastero-Gil:2018uep}. Moreover, the information of the coupling between the inflaton and the radiation in the heat bath leading to a growing mode is contained in the function $G(Q_N)$ and it reads \cite{Benetti:2016jhf}, 
\begin{eqnarray}
G(Q) = 1 + 0.335\,Q^{1.364} + 0.0185\,Q^{2.315}\,.
\label{growing}
\end{eqnarray}
In addition, we have used the relation $\rho_r/V(\phi) = \epsilon\,Q/2(1+Q)^2$ and the approximation of the thermalized inflaton fluctuation, $1+ 2\,n_N \simeq 2\,T_N/H_N $ and $T_N/H_N = 3\,Q_N/C_T$ in order to get the last line in Eq.(\ref{spectrum}) as Ref.\cite{Bastero-Gil:2018uep}. By using Eqs.(\ref{spectrum},\ref{growing}), furthermore, the scalar spectral index is determined as
\begin{eqnarray}
n_s &=& 1 + \frac{d\ln \Delta_{\mathcal{R}}}{dN}
\nonumber\\
&=& 1 + \frac{Q_N}{3+5\,Q_N}\frac{\big(6\,\epsilon - 2\,\eta \big)}{\Delta_{\mathcal{R}}}\,\frac{d\Delta_{\mathcal{R}}}{dQ_N}\,,
\label{ns-grow}\\
\frac{d\Delta_{\mathcal{R}}}{dQ_N} &=& \frac{5\,C_T^3}{12\,\pi^4\,g_R}\Bigg[\left(1+ \frac{\sqrt{3}\,\pi\,Q_N}{\sqrt{3+4\pi\,Q_N}}\right)\frac{\left(0.457\,Q^{0.364}+0.0428\,Q^{1.315}\right)}{Q_N^2}
\nonumber\\
&-& \left(2+ \frac{\sqrt{3}\,\pi\,Q_N}{ \sqrt{3+4 \pi\,Q_N}}+\frac{2 \,\sqrt{3}\, \pi^2\,Q_N^2}{(3 + 4\,\pi\, Q_N)^{\frac32}}\right)\frac{\big( 1 + 0.335\,Q^{1.364} + 0.0185\,Q^{2.315} \big)}{Q_N^3} \Bigg].\nonumber
\end{eqnarray}
While the tensor-to-scalar perturbation ratio, $r$ is obtained by the following formula
\begin{eqnarray}
r = \frac{\Delta_T}{\Delta_{\mathcal{R}}} = 16\,\epsilon\left[ \frac{6\,Q_N^3}{C_T}\left(1+ \frac{\sqrt{3}\,\pi\,Q_N}{\sqrt{3+4\pi\,Q_N}}\right)G(Q_N)\right]^{-1}
\label{tensor-scalar-grow}
\end{eqnarray}
where $\Delta_T$ is the power spectrum of the tensor perturbation and we have used $\Delta_T = 2H^2/\pi^2M_p^2 = 2U(\chi)/3\pi^2 M_p^4$  which is the same form as in the standard (cold) inflation result for the primordial gravitational waves. 

%%%%%%%%%%%%%%%%%%%%%%%%%%%%%%%%%%%%%%
\section{Confrontation with the data}
\label{confront-data}
%%%%%%%%%%%%%%%%%%%%%%%%%%%%%%%%%%%%%%
\begin{figure}[!h]	
	\includegraphics[width=12cm]{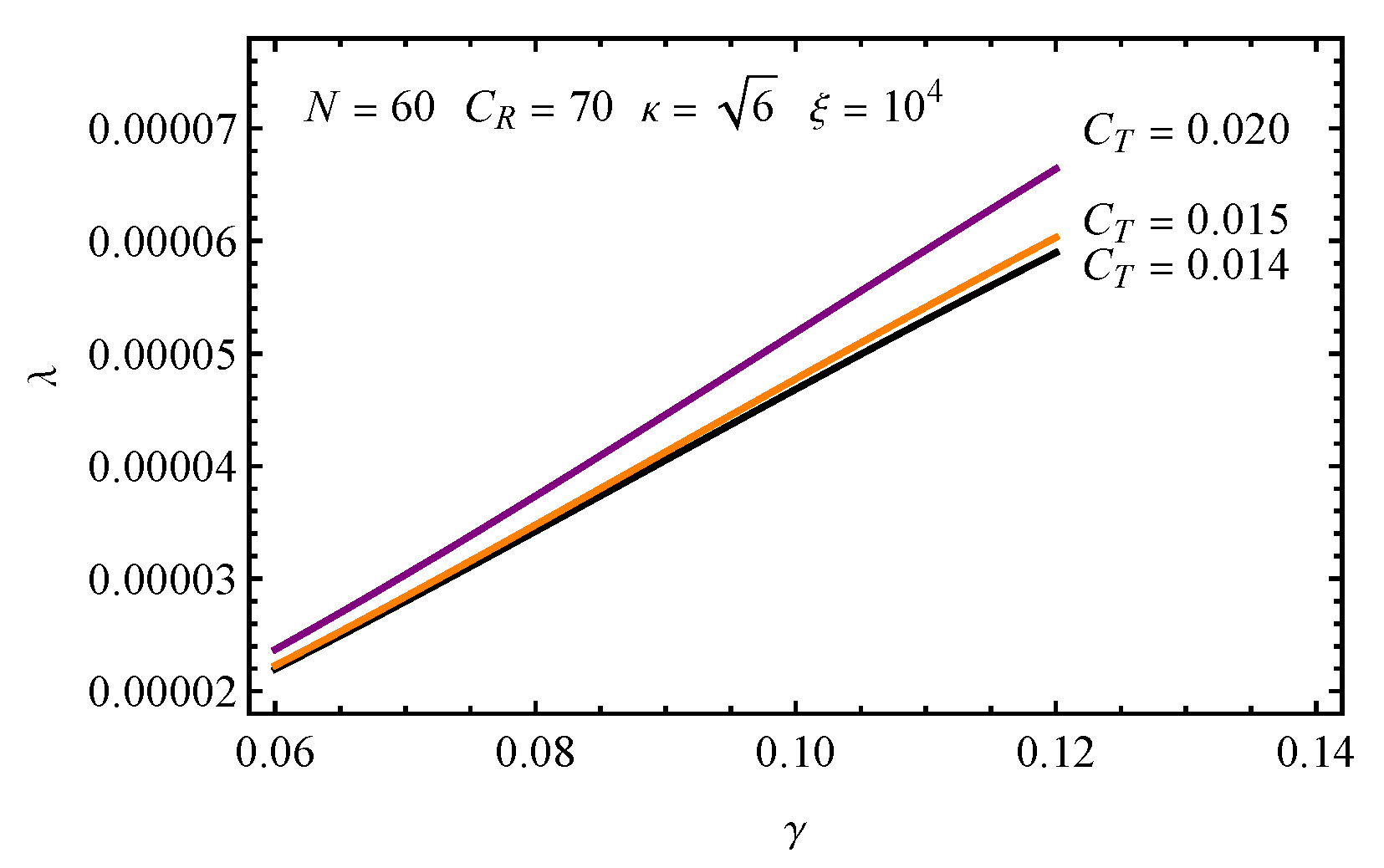}
	\centering
	\caption{We display $\lambda$ as a function of $\gamma$ obtained from Eq.(\ref{lambda-cobe-normalize}) for $M_{p}=10\Lambda,\,\xi=10^{4}$, $C_{R}=70,\,\kappa\approx \sqrt{6}$ and $N=60$.  As $\gamma$ increases, the magnitude of $\lambda$ needed to produce the
correct amount of scalar perturbations also increases.}
	\label{stplot111}
\end{figure}
%%%%%%%%%%%%%%%%%%%%%%%%%%%%%%%%%%%%%%%%%%%%%%%%%%%%%%%%%%%%%%%%%%%%%%%%%%%%%%%%%%%%%%%
\begin{table}[h!]
\begin{center}
\caption{We show a set of parameters $(C_{T},\,\gamma,\,\lambda)$ obtained from Eq.(\ref{lambda-cobe-normalize}) in which their values are constrained by the COBE normalization condition given in Eq.(\ref{1r-strong}). Here we have used various values of $C_{T}$ in order to obtain viable values of $\gamma,\,\lambda$ and applied $M_{p}=10\Lambda$, $\xi=10^{4}$, $C_{R}=70$, $\kappa\approx \sqrt{6}$ and $N=60$.}
\label{tab}
\begin{tabular}{c|c|c}
\hline\hline
\quad $C_{T}$ \quad\quad & \quad $\gamma \times 10^{-2}$ \quad\quad & \quad $\lambda\times 10^{-5}$ \quad\quad \\
\hline
\multirow{4}{*}{0.014} & 7.00 & 2.80\\
& 8.50 & 3.74\\
& 10.00 & 4.68\\
& 11.50 & 5.60\\
\hline
\multirow{4}{*}{0.015} & 7.00 & 2.41\\
& 8.50 & 3.80\\
& 10.00 & 4.78\\
& 11.50 & 5.73\\
\hline
\multirow{4}{*}{0.020} & 7.00 & 3.04\\
& 8.50 & 4.09\\
& 10.00 & 5.19\\
& 11.50 & 6.28\\
\hline
\end{tabular}
\end{center}
\end{table}
%%%%%%%%%%%%%%%%%%%%%%%%%%%%%%%%%%%%%%%%%%%%%%%%%%%%%%%%%%%%%%%%%%%%%%%%%%%%%%%%%%%%%%%%%
In this section, we will constrain the inflation potential with the COBE normalization condition \cite{Bezrukov:2008ut} to fix the parameters in the non-minimal warm inflation with the quantum corrected self-interacting potential. According to the Planck 2018 data, the inflaton potential must be normalized by the slow-roll parameter, $\epsilon$ and satisfied the following relation at the horizon crossing $\phi=\phi_{N}$ in order to generate the observed amplitude of the cosmological density perturbation ($A_{s}$):
\begin{eqnarray}
\frac{U(\phi_N)}{\epsilon(\phi_N)} \simeq (0.0276\,M_{p})^{4}\,.
\label{1r-strong}
\end{eqnarray}
Having used the potential in Eq.(\ref{U-chi}) and the slow-roll $\epsilon$ parameter in Eq.(\ref{SR-epsilon}), we find
\begin{eqnarray}
\left(\frac{M_p}{\Lambda}\right)^2\left(1 + \gamma\,\frac{\xi\,\phi_N^2}{M_p^2} \right) = \frac{\sqrt{3\,\lambda}}{2\,(0.0276)^{2}}\left(\frac{\phi_N}{\Lambda}\right)^{2(1+\gamma)}\,.
\label{lambda-cobe-normalize}
\end{eqnarray}
The resulting constraint is plotted in Fig.\ref{stplot111} by using the definition of $\phi_N$ in Eq.(\ref{Q-N}). The magnitude of $\lambda$ needed to produce the observed amplitude of scalar perturbations increases linearly for increasing $\gamma$. For reference, we consider various values of $C_T$ and figure out a pair of $(\lambda,\,\gamma)$ in which their values produce the observed amplitude of scalar perturbations. More interestingly, the numerical values of the self-interacting coupling, $\lambda$, shown in Fig.\ref{stplot111} are consistent with the results of the running coupling $\lambda$ up to two-loop corrections in the standard model of particle physics that is very close to zero at the GUT scale which would be a typical scale of inflation \cite{Degrassi:2012ry}. In addition, the values of the $\lambda$ coupling is in order $\lambda \sim 10^{-5}$ and still are much bigger than the unnaturally small of $\lambda \sim 10^{-13}$ for the minimal coupling cold inflation \cite{Guth:1982ec,Hamada:2014iga}.  

We present the COBE constrained results for relations between the $\lambda$ and $\gamma$ parameters with varying values of the $C_T$ in Table \ref{tab} and then compare the predictions in the $(r-n_{s})$ plane of the latest Planck 2018 data by using the expressions of the $n_s$ and $r$ in Eqs.(\ref{ns-grow}) and (\ref{tensor-scalar-grow}), respectively with the growing mode in (\ref{growing}).

\begin{figure}[!h]	
	\includegraphics[width=12cm]{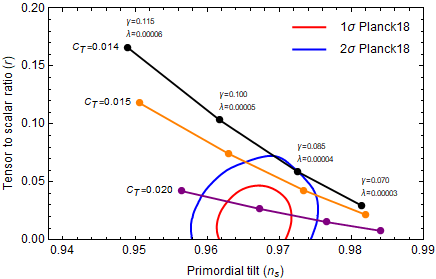}
	\centering
	\caption{We compare the theoretical predictions of the strong limit $Q>1$ including the growing mode effects Eqs.(\ref{ns-grow}) \& (\ref{tensor-scalar-grow}) for $C_{T}=0.020$ (purple), $C_{T}=0.015$ (orange) and $C_{T}=0.014$ (black) in the $(r-n_{s})$ plane for various values of $\gamma$ and $\lambda$ given in Table (\ref{tab}) constrained by the COBE renormalization condition by using $C_{R}=70$, $\xi=10^{4}$, $N=60$ and $M_{p}=10\Lambda$ with Planck’18 results for TT, TE, EE, +lowE+lensing+BK15+BAO.}
	\label{stplot1111}
\end{figure}
\begin{figure}	
	\includegraphics[width=11cm]{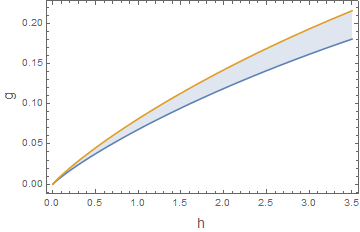}
	\centering
	\caption{The allowed region of the possible values of the $g$ and $h$ from Eq.(\ref{C11}) due to the range of $C_T$ for $0.014 \lesssim C_T \lesssim 0.02$ where $g$ and $h$ are the Yukawa couplings of the inflaton-heavy fermions and the heavy fermions-light singlet scalar and fermion fields, in the supersymmetric model respectively.  }
	\label{gvsh}
\end{figure}

From Fig.\ref{stplot1111}, we present the confidence contours in the $(n_{s},\,r)$ plane. The value of $C_T$ is varied for each trajectory. The curves in this figure are related to $C_T$ as: $0.014$ (black), $0.015$ (orange) and $0.020$ (purple) from the bottom curve to the top one. With a set of input parameters when $C_T$ decreases the curve is shifted upward. The proper set of the parameters $C_{R}=70$, $\xi=10^{4}$, $N=60$ and $M_{p}=10\Lambda$ is used. With these values of the parameters, we find that in order to fit inside the $2\,\sigma$ confidence level of the Planck 2018 data, the range of $C_T$ is in $0.014 \lesssim C_T \lesssim 0.02$ and it dose not exceed the upper bound $0.020$ from the constraint of the power spectrum \cite{Bastero-Gil:2018uep}. More importantly, the given range of the parameter $C_T$, $0.014 \lesssim C_T \lesssim 0.02$ can consequently provide the possible values of the couplings $g$ and $h$ that are encoded in the $C_T$ as shown in Eq.(\ref{C11}). The allowed region in the parameter space of the $g$ and $h$ is depicted in Fig.\ref{gvsh}. To make consistent results between theory and observation, in addition, this requires that the cut-off, $\Lambda$ of the inflaton field should be less than the Planck mass around one order of magnitude in contrast to cold inflation that usually imposes $\Lambda \sim M_p$. 

%%%%%%%%%%%%%%%%%%%%%%%%
\section{Conclusion}
\label{conclusion}
%%%%%%%%%%%%%%%%%%%%%%%%
In this work, we presented the theoretical study of the non-minimal coupling warm inflation with the quantum-corrected self-interacting inflaton potential. The slow-roll dynamics of warm inflation in the Einstein frame is analyzed by using the dissipative coefficient as linear function of temperature. At the large field approximation in warm inflation, the universal bound for the quantum-corrected parameter, $\gamma$ is modified by the dissipative coefficient. With the proper set of the parameters, the universal bound of warm inflation is bigger than that of cold inflation. This indicates that the inflaton field in warm inflation is smaller that of the cold one at the end of inflation. Having used the COBE normalization of the observed amplitude, we found that the relationship between the self-interacting coupling, $\lambda$ and the quantum-corrected parameter, $\gamma$ is linear and the value of the $\lambda$ is in order $\mathcal{O}(\lambda) \sim 10^{-5}$ for $0.06< \gamma < 0.1$. The constraint of the $\lambda$ coupling from COBE is consistent with the renormalization group result at the GUT scale. We continuously compared the tensor to scalar ratio ($r$) and spectral index ($n_s$) from the theoretical results to the Planck 2018 data. As results, the given sets of the model's parameters provide good agreement with the Planck 2018 observational data. To make the theoretical results locating inside the 2$\sigma$ confidence level, it was found that the range of the parameter from the dissipative coefficient, $C_T$ is in range $0.014\lesssim C_T \lesssim 0.02$ and the lower bound of the $C_T$ parameter is constrained in this work. Consequently, we have also used the range of $C_T$ to evaluate the allowed region in the parameter space $g$ and $h$ in terms of the supersymmetric model that are used to calculate the dissipative coefficient. In addition, the self-interacting coupling, $\lambda$ should be very small and this is consistent with the constraint from COBE. More importantly, in contrast to cold inflation scenario, the cut-off scale of the inflaton, $\Lambda$ is smaller than that of the Planck scale of one order of magnitude to obtain the results compatible with the data. Furthermore, higher order quantum-correction and other form of the inflaton potential are worth for extensively study. More information and accurate observational data might provide more details about the quantum-correction of the inflaton and validity of the warm inflationary universe, especially the observation data of the primordial tensor modes.  %say somethings more a few sentences for more detail and accurate results are settled

%%%%%%%%%%%%%%%%%%%%%%%%%%%%%%%%%%%%%%%%%%
\begin{acknowledgments}
D. Samart is financially supported by the Mid-Career Research Grant 2021 from National Research Council of Thailand under a contract No. N41A640145. P. Ma-adlerd is supported by National Astronomical Research Institute of Thailand (NARIT). P. Channuie acknowledged the Mid-Career Research Grant 2020 from National Research Council of Thailand under a contract No. NFS6400117.
\end{acknowledgments}

\bibliography{warminflation}

%\appendix
%\input{section/Appendix}

\end{document}